\documentclass[12pt]{article}

\usepackage{amssymb}
\def\labelmark{}
\def\void{}

{\ifx\void\labelname\def\junk{\end{displaymath}}
\else\def\junk{\end{eqnarray}}\fi\junk\labelmark\def\labelname{}}

\newcommand{\bra}{\begin{array}}
\newcommand{\era}{\end{array}}
\newcommand{\beq}{\begin{equation}}
\newcommand{\eeq}{\end{equation}}
\newcommand{\bqn}{\begin{eqnarray}}
\newcommand{\eqn}{\end{eqnarray}}

\font\mybb=msbm10  at 12pt
\def\bb#1{\hbox{\mybb#1}}

\font\mybbi=msbm10  at 9pt
\def\bbi#1{\hbox{\mybbi#1}}

\def\BC{\bb C}
\def\_\BC{\bbi C}

\newcommand{\om}{\omega}

\newcommand{\la}{\lambda}
\newcommand{\si}{\sigma}

\newcommand{\be}{\beta}

\newcommand{\te}{\theta}

\newcommand{\pa}{\partial}
\newcommand{\al}{\alpha}

\newcommand{\de}{\delta}
\newcommand{\lga}{\longrightarrow}

\newcommand{\st}{\star}

\newcommand{\da}{\dagger}

\newcommand{\ov}{\over}
\newcommand{\hb}{\hbar}

\newcommand{\sq}{\sqrt}

\newcommand{\ev}{\equiv}
\newcommand{\lb}{\label}

\newcommand{\NP}[1]{ {\it Nucl.~Phys.} {\bf #1}}
 
\newcommand{\PL}[1]{ {\it Phys.~Lett.} {\bf #1}}

\newcommand{\PR}[1]{ {\it Phys.~Rev.} {\bf #1}}
\newcommand{\PRL}[1]{ {\it Phys.~Rev.~Lett.} {\bf #1}}

\newcommand{\MPL}[1]{ {\it Mod.~Phys.~Lett.} {\bf #1}}

\newcommand{\JP}[1]{ {\it J.~Phys.} {\bf #1}:\  Math.~Gen.~}

\newcommand{\AP}[1]{ {\it Ann.~Phys. (N.Y.)} {\bf #1}}
\newcommand{\JMP}[1]{ {\it J. Math.~Phys.} {\bf #1}}

\begin{document}
\begin{titlepage}
\setcounter{page}{1}
\renewcommand{\thefootnote}{\fnsymbol{footnote}}

\begin{flushright}
hep-th/0201203
\end{flushright}

\vspace{6mm}
\begin{center}

{\Large\bf Second Virial Coefficient for Noncommutative Space}

\vspace{12mm}
{\large\bf Ahmed Jellal$^{1,2}$
\footnote{E-mail: {\textsf jellal@gursey.gov.tr }}}
and
{\large\bf Hendrik B. Geyer$^{2}$ 
\footnote{E-mail: {\textsf hbg@sun.ac.za}}}\\

\vspace{6mm}

$^{1}$ {\em Institut f\"ur Physik, 
Technische Universit\"at Chemnitz\\
D-09107 Chemnitz, Germany}\\
$^{2}${\em Institute for Theoretical Physics, University of
Stellenbosch\\
Private Bag X1, Matieland 7602, South Africa}\\
\end{center}

\vspace{5mm}
\begin{abstract}
The second virial coefficient $B_{2}^{\rm nc}(T)$ for  
non-interacting particles moving in a two-dimensional
noncommutative space and in the presence of
a uniform magnetic field $\vec B$ is presented. 
The noncommutativity parameter $\te$ can be chosen
such that the $B_{2}^{\rm nc}(T)$  can be interpreted
as the second virial coefficient for anyons of statistics $\al$
in the presence of $\vec B$ and
living on the commuting plane.  
In particular, 
in the high temperature limit $\be\lga 0$, 
we establish a relation between the parameter $\te$ and 
the statistics parameter $\al$.
Moreover, $B_{2}^{\rm nc}(T)$ can also be interpreted 
in terms of composite fermions.
\end{abstract}
\end{titlepage}

\newpage

\section{Introduction}

One of the most interesting features of two-dimensional systems
of charged particles is their exotic or fractional 
statistics~\cite{wil,lei}. Particles with this
property are known
as anyons and represent an interpolation between bosons and fermions.
In fact, quasiparticles of the fractional quantum Hall 
effect~\cite{aro1}
and solitons of a nonlinear $\si$-model~\cite{zee} 
are two important
candidates for a realization of anyons, since 
they exhibit fractional statistics.

A thermodynamic way to capture and  exhibit fractional 
statistics of two-dimensional
systems of charged particles~\cite{joh} is the so-called 
second virial coefficient 
$B_2(T)$~\cite{osb}. $B_2(T)$ represents
a purely quantum mechanical effect
and appears as the first correction to the ideal gas equation
of state
\beq
\lb{svc}
P/Nk_BT = 1 + B_2(T) N + O(N^2), 
\eeq
where $N$ denotes the particle density. At the classical level this
coefficient is missing for non-interacting particles. Moreover,
$ B_2(T)$ is negative for free bosons,
which means that the pressure $P$ decreases from the classical
value at fixed $N$ and temperature $T$. This effect is
as a consequence of the tendency of bosons  
to overlap.  However, $ B_2(T)$ is positive
for non-interacting fermions as a result of the Pauli exclusion 
principle.

We proceed to study the second
virial coefficient and associated statistics for  
non-interacting particles in
an external uniform magnetic field and 
moving in a two-dimensional noncommutative
space. In fact, we will show
that a description of the fractional statistics 
in terms of noncommutative
geometry~\cite{con} is possible. Basically 
our aim is to clarify the role that
noncommutativity can play in the
statistics of particles. This work follows our
previous investigations
of the present system, where  
interesting phenomena like nonextensive statistics~\cite{jel1},
orbital magnetism~\cite{jel2} and the Hall effect~\cite{jel3} are discussed.

In section 2, we give the energy spectrum and
eigenfunctions of a particle moving on a 
two-dimensional noncommutative plane 
and exposed to a uniform external magnetic field. 
In section~3, after recalling the definition of the
second virial coefficient, we compute its noncommutative
expression. This can be interpreted as the usual second virial coefficient
in terms of an effective magnetic field. 
In section 4, by making a specific choice of the
parameter $\te$, we offer an interpretation of $B_2^{\rm nc}(\be)$ as 
the second virial coefficient for  
anyons in the presence of $\vec B$ and living on the commuting
plane. Moreover, in the high temperature limit,
a relation between the parameter $\te$ and 
the statistics $\al$ is obtained.
In the final section we suggest another interpretation of  
$B_2^{\rm nc}(\be)$ in terms of composite fermions.

\section{Particle in a noncommutative space}

Let us consider a non-interacting particle 
moving in a two-dimensional space~$(x,y)$ under 
the influence of a perpendicular
uniform magnetic field~${\vec B}$. 
In the symmetric gauge, this system is 
described by the Hamiltonian
\beq
\label{sh}
H = \frac{1}{2m}\left[
\left( {p}_x -\frac{eB}{2c} y \right)^2 +
\left( {p}_y +\frac{eB}{2c} x \right)^2 \right].
\eeq
We would like to study (\ref{sh}) on a noncommutative 
plane. For that purpose we can accordingly assume
that the coordinates of the plane are 
noncommuting,
\beq
\label{com}
[\mathbf{x},\mathbf{y}]=i\te,
\eeq
where the parameter $\te$ is a real constant.
Noncommutativity can also be imposed
by treating the coordinates as
commuting, but requiring that the composition of their functions
be given in terms of
the star product
\beq
\label{star}
\st\equiv\exp \frac{i\te}{2} \Big(
{\stackrel\leftarrow\pa}_{x} {\stackrel\rightarrow\pa}_{y}
-{\stackrel\leftarrow\pa}_{y}{\stackrel\rightarrow\pa}_{x}
\Big). 
\eeq
Following this route, we deal with commutative 
coordinates $x$ and $y$, but
replace the ordinary products  with the star product
(\ref{star}). For example,   
the commutator (\ref{com}) is replaced by the expression
\beq
\lb{nc}
x\st y-y\st x=i\te .
\eeq
As usual, canonical quantization of this system is achieved by
introducing
the coordinate and momentum operators ${x}_i,\ {p}_i$ satisfying
\beq
\lb{pm}
[x_{i},p_{j}]=i\hb\de_{ij},\qquad [p_{i},p_{j}]=0,\
\eeq
where $x_1=x$ and $x_2=y$. Thus, we treat our system 
in the framework of the algebra generated
by the commutation relations~(\ref{nc})-(\ref{pm}), 
which implies that all 
subsequent products are replaced by their 
star product counterparts as envisaged 
above. According to 
this prescription, the above Hamiltonian acts
on an arbitrary function $\Psi(\vec{r},t)$ as~\cite{mez}
\beq
\label{nch1}
\bra{l}
H \st \Psi (\vec{r},t) =
\frac{1}{2m}\left[
\left( p_x -\frac{eB}{2c} y \right)^2 +
\left( p_y +\frac{eB}{2c} x \right)^2 \right]
\st\Psi (\vec{r},t)\\
\qquad\qquad\;\;\;\; \equiv H^{\rm nc} \Psi (\vec{r},t).
\era
\eeq
Therefore, the noncommutative version of (\ref{sh}) 
can be inferred to be
\beq
\label{nch2}
H^{\rm nc} = \frac{1}{2m}\left[
\left( {\hat p}_x -\frac{eB}{2c} y \right)^2 +
\left( {\hat p}_y +\frac{eB}{2c} x \right)^2 \right],
\eeq
where the momentum operator ${\hat p}_{i}$ is a $\te$-dependent function 
\beq
\lb{mo}
{\hat p}_{i}=(1-\te l^{-2})p_{i}, 
\eeq
with $l=2l_0$, $l_0=\sq{\hb c\ov eB}$ being the magnetic length. Notice that 
the standard Hamiltonian is recovered in the limit $\te=0$.

The eigenvalue problem
\beq   
\lb{ep1}
H^{\rm nc} \Psi^{\rm nc} = E^{\rm nc} \Psi^{\rm nc},
\eeq
can be solved by introducing a set of
creation and annihilation operators on the complex plane $(z,\bar{z})$
such that
\beq
\lb{ob}
\bra{l} 
{\hat a}^\da =-2i\hat{p}_{\bar z}+{m\om_c\ov 2}{z},\\
{\hat a} =2i\hat{p}_{z}+{m\om_c\ov 2}{{\bar z}},
\era
\eeq
where $\om_c={eB\ov mc}$ is the cyclotron frequency.
These operators
satisfy the commutation relation
\beq
\lb{cr}
[{\hat a}, {\hat a}^{\da}] =2m\hb{\hat\om_c},\qquad 
\eeq
with ${\hat \om}_c=\om_c(1-\te l^{-2})$. Now $H^{\rm nc}$
can be expressed in terms of ${\hat a}$ and ${\hat a}^{\da}$ 
\beq
\label{nch3}
H^{\rm nc}= {1\ov 4m}({\hat a}^{\da}{\hat a}
+{\hat a}{\hat a}^{\da}).
\eeq
To distinguish between the degenerate eigenstates of (\ref{nch3}), 
we define another set of creation and annihilation operators
\beq
\lb{ob}
\bra{l}
{\hat b}^{\da} = 2i \hat{p}_{z} - {m\om_c\ov 2}{\bar z},\\                          
{\hat b} = -2i\hat{p}_{\bar z}-{m\om_c\ov 2} z.                           
\era
\eeq
They both commute with ${\hat a}$ and ${\hat a}^{\da}$ and 
their commutator is again
\beq
[{\hat b}, {\hat b}^{\da}] =2m\hb{\hat\om_c}.
\eeq
Therefore, we can show that the energy spectrum and the
eigenstates are  
\beq
\lb{evs}
\bra{l} 
\Psi_{(n,k)}^{\rm nc}\ev |n,k,\te>=
\sq{1\ov (2m\hb {\hat\om}_c)^{n+k} n!k!}
({\hat a}^{\da})^n ({\hat b}^{\da})^k |0,0>,\\
E_n^{\rm nc}= {\hb {\hat\om}_c} (n+{1\ov 2}).
\era
\eeq
Some remarks are in order at this stage.
The eigenstates are labeled not only by the quantum numbers
associated with  ${\hat a}$ and ${\hat a}^{\da}$, but also 
depend on
an additional degree of freedom.
Moreover, from the expression of the eigenstates we distinguish
three different cases:
(i)- $\te <l^{2}$,
(ii)- $\te >l^{2}$ and
(iii)- $\te =l^{2}$.
The first two cases can be dealt with simultaneously, by
simply considering the replacement $\te\longleftrightarrow  -\te$.
Furthermore, in the subsequent analysis we dot not consider
the case (iii) which would simply lead to a static
critical point not relevant to our discussion.

\section{Noncommutative second virial coefficient}

The second virial coefficient for particles
has been investigated on many occasions, in
particular for
anyons in the absence~\cite{aro3,com} as well as
in the presence of a magnetic field~\cite{joh,aro2}. 
Further study can be found in~\cite{myung}. 
For instance
in~\cite{joh} it is found that $ B_2(T)$ becomes a function
of the fractional statistics $\al$. Now
let us recall the basic definition of $ B_2(T)$ in two dimensions~\cite{joh}:
\beq
\lb{svc0}
B_2(T) = \lim_{S\lga\infty} S\; [{1\ov 2}-{Z_2\ov Z_1^2}],
\eeq
where $S$ is the area of the system and $Z_1$ and $Z_2$ 
are the single-particle and two-particle partition functions,
respectively
\beq
\lb{pf}
Z_i= {\rm Tr} \exp(-\be H_i), \qquad \be={1\ov k_BT}, \qquad i=1,2.
\eeq

We now investigate the second virial coefficient for non-interacting
particles moving on a noncommutative plane and in 
the presence of a
magnetic field. This latter can be defined in the standard
way as (\ref{svc0})
\beq
\lb{nsvc0}
B_2^{\rm nc}(T) = \lim_{S\lga\infty} S\; 
[{1\ov 2}-{Z_2^{\rm nc}\ov (Z_1^2)^{\rm nc}}],
\eeq
where the partition function in the noncommutative
coordinates is
\beq
\lb{npf1}
Z_i^{\rm nc}= {\rm Tr} \exp(-\be H_i^{\rm nc}).
\eeq
Thus, for two particles of energy 
$E_{(n_1,n_2)}^{\rm nc}={\hb{\hat\om}_c}(n_1 + n_2 + 1)$, 
we have
\beq
\lb{npf2}
Z_{2}^{\rm nc}(\be)= \sum_{n_1,n_2} 
\exp(-\be E_{(n_1,n_2)}^{\rm nc}). 
\eeq
The calculation of (\ref{npf2}) depends on what
kind of particles we have. For non-interacting bosons
in the presence of a magnetic field and living on
a noncommutative plane, $Z_{2}^{\rm nc}(\be)$ is 
\beq
\lb{npf3}
Z_{2}^{\rm nc}(\be)= {1\ov 2}\;
\Big([Z_{1}^{\rm nc}(\be)]^2 + 
Z_{1}^{\rm nc}(2\be)\Big).
\eeq
To obtain (\ref{npf3}), we sum independently 
on $n_1$ and $n_2$,
also we add to this sum the contribution 
of the diagonal states for
which $n_1=n_2$. Actually,
the noncommutative second virial coefficient 
$B_{2}^{\rm nc}(T)$ becomes
\beq
\lb{nsvc1}
B_{2}^{\rm nc}(T) = -{1\ov 2}\; \lim_{S\lga\infty}\; S\; 
{Z_{1}^{\rm nc}(2\be)\ov [Z_{1}^{\rm nc}(\be)]^2}.
\eeq
To evaluate explicitly the single-particle 
partition function
$Z_1^{\rm nc}(\be)$, we introduce 
coherent states corresponding to 
the present system 
\beq
\lb{cs}
|\eta,\mu> = e^{-{1\ov 2 {\hat l}^2}(|\eta|^2 + |\mu|^2 )} 
\exp({\eta {\hat a}^{\da}\ov 2i\hb} + 
{\mu {\hat b}^{\da}\ov 2i\hb}) |0,0>.
\eeq
They satisfy the relations
\beq
\lb{ac}
\bra{l}
{\hat a} |\eta,\mu> = {\hb\ov i {\hat l}^2} \eta  |\eta,\mu>,\\
{\hat b} |\eta,\mu> = {\hb\ov i {\hat l}^2} \mu |\eta,\mu>.
\era
\eeq
Now $Z_1^{\rm nc}(\be)$ can be written as follows
\beq
\lb{1ncpf}
Z_1^{\rm nc}(\be) = {4e^{-\be {\hb{\hat\om_c}\ov 2}}
\ov \pi^2 {\hat l}^4}\
\int\; d^2 {\eta} d^2 {\mu}\;
<\eta,\mu| e^{-{\be \ov 2m}{\hat a}^{\da}{\hat a}} 
|\eta,\mu>.
\eeq 
Using the boson-operator identity \cite{lou}
\beq
\lb{boi}
e^{\xi {a}^{\da}{a}}=
\sum_{n=0}^{\infty}{(e^{\xi}-1)^n\over n!}
{a}^{\da n}{ a^n},
\eeq
which holds for any operators ${a}^{\da}$ and $a$ 
satisfying the commutation relation $[{a},a^{\da}]=1$, 
we can show that (\ref{1ncpf}) simplifies
to
\beq
\lb{2ncpf}
Z_{1}^{\rm nc}(\be)= {4e^{-\be {\hb{\hat\om_c}\ov 2}}\ov 
\pi^2 {\hat l}^4}
\int d^2 {\eta} d^2 {\mu}
e^{-2{|{\eta }|^2\ov {\hat l}^2} 
(1-e^{-\be\hb{\hat\om_c}})}.
\eeq
For a system of area $S$, the integral can be calculated 
to be~\cite{cah}
\beq
\lb{pf4}
Z_{1}^{\rm nc}(\be) =
{m{\hat\om}_c S\ov 4\pi\hb}\; 
{1\ov  \sinh\Big({\be\hbar{\hat\om_c}\over 2}\Big)}.
\eeq
Now, from (\ref{nsvc1}) and (\ref{pf4}) we obtain
\beq   
\lb{nsvc2}
B_{2}^{\rm nc}(T) = -{\pi\hb\ov m{\hat\om_c}} 
\tanh\Big({\be\hbar{\hat\om_c}\over 2}\Big).
\eeq
This equation shows the dependence of 
$B_{2}^{\rm nc}(T)$
on the noncommutativity parameter $\te$. The free
fermion case can be worked out in the same way
as above by taking the Pauli exclusion principle
into account,
\beq
\lb{ncfsvc}
B_{2}^{\rm nc}(T)|_{f}=-B_{2}^{\rm nc}(T)|_{b}=
{\pi\hb\ov m{\hat\om_c}}    
\tanh\Big({\be\hbar{\hat\om_c}\over 2}\Big), 
\eeq
where $f$ and $b$ refer to fermions and bosons, respectively.
Notice that the two expressions only differ by a minus
sign. We would like to emphasize  
that the standard results  are
recovered if the noncommutativity parameter
$\te$ is switched off:
\beq
\lb{sr}
B_{2}(T)|_{f} = - B_{2}(T)|_{b} = 
{\pi\hb\ov m{\om_c}}
\tanh\Big({\be\hbar{\om_c}\over 2}\Big).   
\eeq
Comparison of
(\ref{nsvc2}) with (\ref{sr}) suggests that one can interpret the
noncommutative case as a theory of the second virial coefficient on 
a commuting plane with an effective magnetic field
\beq
\lb{ncf}
B_{\rm eff}= B-B\te l^{-2}.
\eeq
Some remarks are in order at this stage. One can measure
the parameter $\te$ in terms of the magnetic length $l^2$:
\beq
\lb{tel}
\te_l={m\ov n}l^2.
\eeq
For $\te>l^2$, ${m\ov n}>1$ and
we obtain a quantized effective magnetic field
\beq
\lb{ncf}
B_{\rm eff}= ({n-m\ov n})B.
\eeq
The quantization can be integer as well as fractional,
depending on whether the ratio ${m\ov n}$ is integer or 
fractional. On the other hand, if we restrict ourselves
to the high temperature limit $(\be\lga 0)$, we find that 
$B_{2}^{\rm nc}(T)$ can be approximated as
\beq
\lb{nsvc3}
B_{2}^{\rm nc}(T) \approx -{\pi l^{2}\ov 8} 
\Big[{x}-{x^3\ov 12} \Big]
+{\pi l^{2}x^3\ov 96}
\Big[\te^2 l^{-4}- 2\te l^{-2} \Big],
\eeq
where $x= {\be\hb\om_c}$. In the same limit, we have for
$B_{2}(T)|_{b}$ 
\beq
\lb{svc2}
B_{2}(T)|_{b} \approx -{\pi l^{2}\ov 96} 
\Big[{x}-{x^3\ov 12} \Big].
\eeq
Therefore, we obtain
\beq
\lb{nsvc4}
B_{2}^{\rm nc}(T) \approx B_{2}(T)|_{b}
+{\pi l^{2}x^3\ov 96}\Big[\te^2 l^{-4}-2\te l^{-2} \Big].
\eeq
From this relation it is clear that in the high temperature 
limit, a correction to the 
standard second virial coefficient is obtained
in terms of the noncommutativity parameter 
$\te$.

\section{Fractional statistics and noncommutativity parameter}

In this section we present an interpretation of 
the second virial coefficient for free bosons moving on
a noncommutative plane. 
For that purpose, let us recall 
a result for the second virial coefficient obtained for
anyons in the presence of a magnetic field~{\cite{joh,aro2}}.  
Based on bosons, Johnson and Canright~{\cite{joh}} found that
the coefficient is given by
\beq
\lb{jc}
B_{(2,b)}^{\al}(T) = -
{\la^2\ov 2x}\tanh\Big({x\over 2}\Big)+{\la^2\ov x}
\Bigg[{e^x\ov \sinh x}(1-e^{-\al x})-\al  \Bigg],
\eeq
where $0\leq\al\leq 1$ and 
$\la=({2\pi\hb^2\ov mkT})^{1\ov 2}$.
Moreover, they showed that the fermion-based result 
can be derived from the bosonic case, leading to
\beq
\lb{af}
B_{(2,f)}^{\al}(T)=B_{(2,b)}^{\al\pm 1}(T).
\eeq

For this we can establish a relation between the noncommutativity
parameter $\te$ and the statistics parameter $\al$. Alternatively, 
we consider the possibility to describe exotic statistics in terms 
of noncommutative geometry. To do this, we can identify 
(\ref{nsvc2}) with (\ref{jc})
\beq
\lb{id1}
{\la^2\ov 2{\hat x}_{\al}}
\tanh\Big({{\hat x}_{\al}\over 2}\Big)=
{\la^2\ov 2x}\tanh\Big({x\over 2}\Big)-
{\la^2\ov x} \Bigg[{e^x\ov \sinh x}(1-e^{-\al x})-\al  \Bigg],
\eeq
where ${\hat x}_{\al}=x(1-\te_{\al}l^{-2})$.
It is clear that this equation cannot be solved 
Explicitly. Some approximations are thus required to
proceed analytically.
Let us consider the high 
temperature limit, where 
$B_{(2,b)}^{\al}(T)$ can be expressed as 
\beq
\lb{jc2}
B_{(2,b)}^{\al}(T) \approx -{\pi l^{2}\ov 8} 
\Big[{x}-{x^3\ov 12} \Big] +
{\pi l^{2}\al\ov 2} \Big[x+{x^2\ov 3} \Big].
\eeq
From equations (\ref{nsvc3}) and (\ref{jc2}), we find that
$\te_{\al}$ satisfies a second order equation
\beq
\lb{id3}
\te_{\al}^2  - 2\te_{\al} l^{2} -
{48l^{4} \al} \Big[{1\ov x^2}+{1\ov 3x} \Big] = 0.
\eeq
According to the condition $\te > l^{2}$, there is only one valid
solution of this equation
\beq
\lb{id4}
\te_{\al}=l^{2}\Big[1+\sq{1+48\al f(x)} \Big],
\eeq
where we have set $f(x)= {1\ov x^2}+{1\ov 3x}$ and
it is always positive when $\be\lga 0$.
Therefore, in the high temperature limit,
and when $\te$ is fixed to be $\te_{\al}$, one can interpret
$B_{2}^{\rm nc}(T)$ as the second virial coefficient
for anyons of statistics $\te_{\al}$. We close this section
by mentioning that for small $\te$, the last equation
becomes
\beq
\lb{id5}
\te_{\al}= {24l^{2}\al} \Big[{1\ov x^2}+
{1\ov 3x}\Big].
\eeq
Clearly the condition $\te > l^{2}$ implies the further constraint 
$24\al \Big[{1\ov x^2}+{1\ov 3x}\Big]>1$.

\section{Concluding remarks}

Complementary to the analysis above, we can also give 
an interpretation of
$B_{2}^{\rm nc}(T)$ in terms of composite fermions 
(CF)~\cite{cf},
which have been introduced as a new type of particle in
condensed matter physics to provide an explanation of the 
fractional quantum Hall effect~\cite{gir}. CF's are particles
carrying an even number $2p$ $(p=1,2,\cdots )$ 
of flux quanta (vortices).
They have the same charge, spin and statistics 
as particles, but they
differ from them since they experience an effective magnetic field   
\beq
\lb{cfm}
B^*=B - 2pN\Phi_0,
\eeq
where $\Phi_0={hc\ov e}$ is the unit of flux.
To interpret $B_{2}^{nc}(T)$ as the second virial coefficient 
for composite fermions we should choose $\theta$ such that
\beq
\lb{emf}
B_{\rm eff}|_{\te=\te_c}=B^{*}.
\eeq
We solve this to obtain
\beq
\lb{tef}
\te_c={4\ov \pi}{pn\ov n_B^2},
\eeq
where $n_B={\Phi_B\ov \Phi_0}$. 
Here $\Phi_B$ is the flux due to $\vec B$
and $n=NS$ is the number of particles. Thus, 
this equation shows the
possibility to express the noncommutativity 
parameter $\te$
in terms of the flux quanta.

In conclusion, non-interacting particles on a  
noncommutative plane and in the presence of a magnetic 
field have been considered.
The second virial coefficient$B_{2}^{\rm nc}(T)$ 
corresponding
to this system is obtained in terms of 
the noncommutativity parameter
$\te$ . By fixing the latter to be $\te_{\al}$,
we interpreted  $B_{2}^{\rm nc}(T)$
as the second virial coefficient for 
anyons of statistics $\al$
moving on the commuting plane and
experiencing a magnetic field. Moreover
in the high temperature
limit, a relation between the
fractional statistics $\al$ and
the parameter $\te$ is obtained. These results illustrate 
the possibility to deal with the particle statistics in 
terms of noncommutative
geometry.


\end{document}